\documentclass[useAMS,usegraphicx,usenatbib]{mn2e}

\title[Globular Cluster Candidates in NGC 891]{Globular Cluster Candidates in NGC 891$^1$}
\author[Harris et al.]{W.~E.~Harris$^1$, M.~Mouhcine$^2$, 
M.~Rejkuba$^3$ R.~Ibata$^4$\\
$^{1}$Department of Physics \& Astronomy, McMaster University,
      Hamilton ON L8S 4M1, Canada \\
$^{2}$Astrophysics Research Institute, Liverpool John 
      Moores University, Twelve Quays House, Egerton 
      Wharf, Birkenhead, CH41 1LD, UK \\
$^{3}$ESO, Karl-Schwarzschild-Strasse 2, 
      D-85748 Garching, Germany \\
$^{4}$Observatoire Astronomique de Strasbourg (UMR 7550),
      11, rue de l'Universit\'e, 67000 Strasbourg, France}

\begin{document}
\maketitle
\label{firstpage}
\begin{abstract}
We use deep images taken with the Advanced Camera for Surveys on board 
the Hubble Space Telescope 
of the disk galaxy NGC~891, to search for globular 
cluster candidates. This galaxy has long been considered to be a 
close analog in size and structure to the Milky Way and is nearly 
edge-on, facilitating studies of its halo population. These 
extraplanar ACS images, originally intended to study the halo
field-star populations, 
reach deep enough to reveal even the faintest globular 
clusters that would be similar to those in the Milky Way. From the three pointings 
we have identified a total of 43 candidates after culling by 
object morphology, magnitude, and colour. We present $(V,I)$ photometry
for all of these, along with measurements of their effective radius
and ellipticity. The 16 highest-rank candidates within the whole sample are
found to fall in very much the same regions of parameter space occupied by 
the classic Milky Way globular clusters.  Our provisional conclusion
from this survey 
is that the total globular cluster population in NGC 891 as a whole
may be almost as large as that of the Milky Way.
\end{abstract}

\begin{keywords}
galaxies:  star clusters -- globular clusters: general
\end{keywords}

\section{Introduction}
\footnotetext[1]{This work was based on observations with the 
NASA/ESA Hubble Space Telescope, obtained at the Space Telescope 
Science Institute, which is operated by the Association of 
Universities for Research in Astronomy, Inc.,under NASA contract 
NAS 5-26555. }

The stellar content, both the diffuse component and star clusters, 
of the outskirts of galaxies are among the oldest and the most 
metal-poor stellar components of galaxies. 
Their properties are clues to the understanding of how galaxies 
have assembled their mass, and constrain the early phases of 
galaxy formation. Recent Hubble Space Telescope (HST) imaging and 
ground-based spectroscopy of star clusters have revolutionized our 
understanding of galaxy formation and evolution. However, elliptical 
and lenticular galaxies have received by far the largest amount 
of attention due to their much richer globular cluster (GC) systems 
and freedom from internal extinction. Conversely, our knowledge 
of GC systems in spirals is still limited to essentially the Galaxy 
and M31, supplemented by a handful of more distant galaxies 
\citep{kisslerpatig1999, goudfrooij2003, chandar2004, rhode2007, 
spitler2006, mora2007}. 
Until very recently, available data indicated that the GC systems 
of the Galaxy and M31 were quite similar and thus it was natural 
to assume this held true for spirals in general. However, in the 
last few years, a growing body of evidence suggests that there are 
some important differences between the GCSs of these two key spirals. 
Among these is the likelihood that M31 possesses young, thin disk 
massive clusters \citep{morrison2004}, as well as intermediate-age 
massive clusters \citep[e.g.][]{beasley2005}, populations that are 
{\it not} present amongst the classically old GCs in the Galaxy. 
The young and intermediate-age massive clusters in M31 are 
significantly more massive than any open clusters in the Milky Way. 
Similar clusters have now been seen in a variety of other spirals 
\citep[e.g. M33;][]{chandar2006}. The GC population in M31 also contains 
a subset of object of extended and diffuse nature, unlike any 
clusters found in the Milky Way \citep{huxor05,huxor08}. 
These objects are found to have similar stellar populations to   
those of the Milky Way's old GCs \citep{mackey06}, and to fill the gap 
in structural-parameter space between GCs and dwarf spheroidals \citep{huxor05}.

NGC 891, a nearby large late-type disk galaxy, has often been 
described in shorthand as a ``clone''  of the Milky Way, since 
it has a very similar total luminosity, bulge size, and disk with 
prominent dust lanes \citep[e.g.][]{vanderkruit1984}. A study of   
surface-brightness photometry \citep{mor97} revealed the presence of 
an extensive thick disk. Its orientation almost precisely edge-on 
to our line of sight makes it particularly attractive for studies 
of its disk and halo stellar populations, facilitating comparative 
studies of components such as the thick disk, stellar streams and 
substructures, and the total visible mass and extent of the halo. 
In addition, NGC 891 is close enough to us that HST imaging is 
easily capable of resolving the halo stars, enabling direct 
star-by-star statistical studies of its old stellar populations. 
Using deep HST imaging of three extra-planar fields extending 
outward to more than 10 kpc from the plane of the galaxy, we have 
been studying the resolved stellar populations to investigate the
spatial strucuture, the bi-dimensional distribution, to search 
for substructure, and to constrain the metallicity distribution 
functions. 

One obvious component of its halo that has not been 
investigated to similar detail is the GC system. If this galaxy 
is indeed similar to the Milky Way, then $\sim 100-200$ of these 
luminous, old star clusters should be present and relatively easy 
to find. A signal pointing in the opposite direction, however, 
is found in the pioneering effort to search for a GC population in 
NGC 891 by \citet{vdb82}, from starcounts on wide-field photographic 
plates. They found no conclusive evidence for any GCs and an upper 
limit of $S_N \la 0.2$ on the specific frequency. It should be 
realized, however, that this imaging material was taken with 
$1''$ seeing quality and had a much brighter limiting magnitude
than is conventionally possible with modern cameras, 
making it difficult to find traces of a GC 
population in the presence of significant field contamination. 
Also, with such material, individual GCs cannot be distinguished either 
from foreground stars or faint, small background galaxies, and both 
types of contaminants
are present in large numbers within the NGC 891 field.

In the present paper, for the first time we conduct a search for, 
and characterization of, individual GCs in NGC 891. Contrary to our 
original expectations, we find that this galaxy does indeed have a 
roughly normal GC population for its size and type. Throughout this 
paper, we adopt the distance modulus $(m-M)_0 = 29.94$ ($d=9.7$ Mpc) 
derived by \citet{Mouhcine07}, along with a foreground reddening 
$E_{V-I} = 0.08$ and $A_V=0.20$ from the NASA/IPAC Extragalactic
Database (NED).

\section{The Database and Initial Searches}

The raw data consist of the same deep HST images taken with the Wide 
Field Channel of the Advanced Camera for Surveys (ACS/WFC) camera used 
in previous papers in this series \citep{Mouhcine07, Ibata2008, 
Rejkuba2008}.  In these papers, \citet{Mouhcine07} discuss the 
stellar halo population; \citet{Ibata2008} 
present a structural analysis to
establish the presence of a thick disk and a stellar halo as well as
small-scale substructures in the halo; and \citet{Rejkuba2008}
investigate the stellar populations of the thick disk and the inner halo.

\begin{figure}
\includegraphics[angle=0,width=0.5\textwidth]{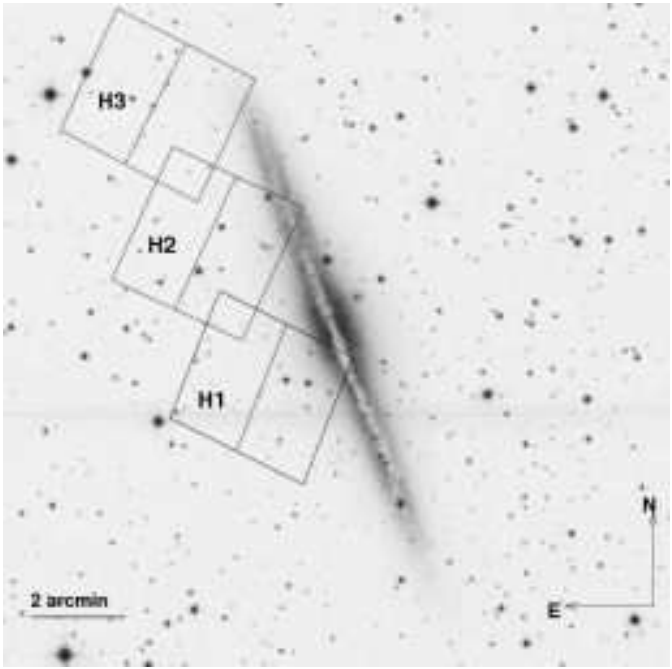}
\caption{Location of the three HST/ACS target fields along the 
northeast side of the disk of NGC 891.  See \citet{Mouhcine07}
for additional description.}
\label{chart}
\end{figure}

Images were taken at three pointings running parallel 
to, and east of, the disk of NGC 891. A finder chart for these three 
fields (labelled H1, H2, H3) is shown as Figure 1 of \citet{Mouhcine07}
and is reproduced here in Figure \ref{chart}.
The three fields were placed to probe both the disk and thick disk 
of the galaxy over a wide radial range as well as the inner halo, 
with slight overlaps between fields to ensure photometric consistency. 
For each field, the ACS/WFC observations consist of three full-orbit 
integrations in both F606W and F814W (for convenience we refer to these 
below as $V$ and $I$). The photometry reaches to ${\rm I\sim 29}$, 
approximately 3~magnitudes below the tip of the red giant branch. 
Artificial-star experiments were performed to estimate the completeness 
of our data set. These were carried out in the usual fashion with  
artificial stars generated from the stellar point spread function (PSF) 
constructed during the photometry measurements. 
The 50 per cent completeness limit, for colours typical of old simple 
stellar populations (i.e., $(V-I)\la 1.3$), occurs at $I\sim 27.3 - 28.2$
depending on galactocentric distance and crowding.
\citet[][]{Rejkuba2008} give a thorough description of the data, the 
data reduction process, and the completeness simulations; we therefore 
refer the reader to this paper for the details. 

To start the process of identifying candidate GCs, we visually
inspected every part of each field and marked any objects that 
might possibly be globular clusters.  The selection was
done independently in each filter.
The western sides of fields H1 and H2 particularly are heavily
contaminated by the bulge light and planar dust lanes in the
galaxy, so to aid the identification process we first subtracted
off median-smoothed versions of the fields (unsharp masking)
using a $39 px-$square median filter box.
The images very obviously resolve the stellar halo red-giant
population of NGC 891, with the brightest normal red giants lying 
at $I\simeq 26$. The faintest known GCs in the Milky Way lie at 
this level or brighter (with the vast majority at $M_I < -6$, 
translating to $I\simeq 24$ at the distance of NGC~891) and so 
these images should sample virtually the entire GC luminosity 
distribution.  The fact that the NGC 891 halo stars are well resolved 
also means that at least some of its GCs could show some degree of resolution 
of their individual red giants outside their cores; some examples 
of these are shown later.

The main search criteria for this initial round of selection were 
that GC candidates should be (a) at least as bright as the brightest 
halo red giants, (b) relatively uncrowded, and (c) morphologically 
\emph{symmetric}; that is, with no features such as small spiral 
arms, tidal tails, companions etc. that would be clear markers of 
faint background galaxies. This initial selection was deliberately 
made very generous; at this stage, no candidate was rejected by 
ellipticity, scale size, location, or colour.

\section{Refining the Candidate List}

With these preliminary lists identified, 
contaminants were weeded out through a series of objective criteria.
First was to match the lists in both filters and keep only those 
identified on both $V$ and $I$.  This step quickly removes any 
extremely red or extremely blue objects, or potential artifacts 
appearing on one filter only.

The next stage was to measure the characteristic size and shape
of each object.  We used the \emph{ISHAPE} profile-fitting code
of \citet{lar99} to  
derive the effective (half-light) radius $r_h$
and ellipticity $e = (1-b/a)$ of each object.  On each image,
15 to 20 moderately bright, uncrowded stars were combined with
\emph{iraf/daophot} to construct an empirical PSF for the frame.  
For each candidate GC, \emph{ISHAPE} was then used to convolve
the PSF with a ``King30'' model profile, i.e. a \cite{kin62} model
with concentration index $r_t/r_c = 30$ characteristic of the average
for known globular clusters.  Here as usual $r_c$ and $r_t$ are
the King-model core and tidal radii.
The assumed effective radius $r_h$ and
axial ratio $b/a$ of the model are then varied till a best fit is
achieved \citep[see][for extensive discussion of the technique]{lar99}.
Extensive simulations by \citet{lar99}
and \citet{har08} show that the derived $(r_h, e)$ are nearly
independent of the assumed $r_t/r_c$ ratio in situations like this
one where $r_h \la fwhm(PSF)$ (see below).

Because NGC 891 is relatively nearby, the size range $r_h \simeq 2 - 5$
parsecs typifying the majority of GCs in the Milky Way converts to
a range of angular size $r_h = 0.04'' - 0.11''$ or about 1 to 2.5 pixels 
on the ACS/WFC camera.  Objects this extended can be termed ``partially
resolved'' because their intrinsic radii are comparable to, or smaller
than, the PSF, but they are easily distinguishable
from stars:  The stellar PSF on the ACS/WFC has a 
FWHM $= 1.9 \pm 0.1$ pixels, and extensive tests of the ISHAPE profile 
fitting code \citep[e.g.][]{lar99,har08} show that the effective radii 
of partially resolved objects such as these can be correctly detected
and measured down to 20\% of the PSF FWHM, and even smaller under conditions
of high S/N.  Thus all GCs with characteristic sizes comparable to
those in the Milky Way should be easily found.  These size measurements
can then quickly be used to eliminate all stars from our sample
of candidate GCs.

Use of the ellipticities is also effective. Small differences
in mean GC shapes have been found from galaxy to galaxy that are not yet
well understood \citep[e.g.][]{har02,han94}, but these differences are minor,
and the vast majority of
known GCs in any galaxies surveyed so far are quite round in projected
shape ($e \la 0.15$) and virtually none are known with $e \ga 0.3$.

\begin{figure}
\includegraphics[angle=0,width=0.5\textwidth]{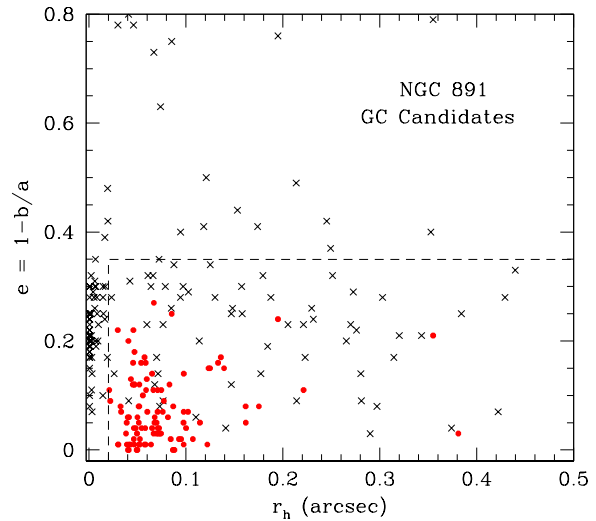}
\caption{Distributions of globular cluster candidates by half-light radius
and eccentricity, as measured by ISHAPE.  Objects in NGC 891 are shown as
crosses, while the Milky Way globular clusters projected to the same distance
are shown as solid red dots.  We reject any candidates lying to the left 
of or above the dashed lines, as being either starlike or too elliptical.}
\label{firstsample}
\end{figure}

In Figure \ref{firstsample} we show the distribution of the candidate
objects measured in both $V$ and $I$ by their \emph{ISHAPE-}determined
values of $r_h$ and $e$.  For comparison, the same data for the Milky 
Way GCs \citep{har96} are shown as well (solid dots in the Figure), where 
their $r_h$ values are projected to the size they would appear if placed at the
9.7-Mpc distance of NGC 891.  The candidate list has a large number
of objects at $r_h < 0.02''$; these are probable stars and can rather
safely be rejected.  We also conservatively eliminate any with $e > 0.35$,
a limit well above the most elliptical known Milky Way clusters.
Additional visual inspection of these highly elongated objects confirms
that they are likely to be background galaxies; none have any indication
of being resolved into stars, and many are next to other obvious galaxies
on the fields.  The one exception we made to this exclusion was for a
few very faint candidates that \emph{may} be resolved into stars and
were also somewhat elongated.  Three examples of these are shown
in one of the figures below.  These have luminosities and effective
radii that, if they are indeed clusters, would make them roughly
comparable to the Palomar-type clusters in the Milky Way or even fainter.
We regard their true nature as more uncertain than most of the rest
of the sample, but keep them in the list.

\begin{figure}
\includegraphics[angle=0,width=0.5\textwidth]{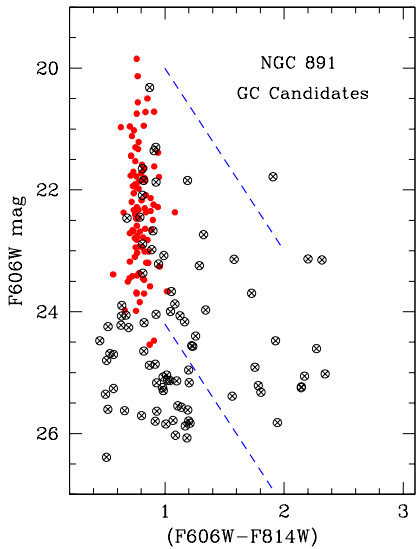}
\caption{Distribution of candidate GCs in the colour-magnitude plane.
The NGC 891 objects shown here (open symbols with crosses) are ones
with effective radii and ellipticities within the range of known
globular clusters.  The Milky Way clusters (solid red dots), projected
to the distance of NGC 891 and with its foreground reddening applied,
are shown for comparison.  The two dashed lines show reddening
trajectories, with slope $A_{606W} = 3 E_{606W-814W}$.
}
\label{cmdraw}
\end{figure}

The third stage of culling is to use colour and magnitude.  The raw
colour-magnitude diagram (CMD) for the 81 candidates surviving the structural
parameter tests is shown in Figure \ref{cmdraw}.  Among these are
five which appear on the small overlap areas between H1/H2 and H2/H3
and were identified on both.  In the CMD,
we plot the apparent magnitudes and colour indices in the 
filter system native to ACS ($F606W, F814W$), 
as measured through \emph{iraf/daophot} and aperture
photometry with an aperture radius $r_{ap}=3$ px $= 0.15''$.
This aperture size corresponds to about $2.5 r_h$ for a median 
GC in the Milky Way and thus safely includes most of its light.
Cluster-to-cluster differences in $r_h$ mean that any fixed-aperture
photometry will not include the same fraction of their true
total magnitude, but at this stage we are more strongly interested
in the cluster colors, and for the many GC candidates projected
on the bulge and disk regions of NGC 891, the aperture radius needs
to be as small as possible to avoid large uncertainties from field contamination.
The calibration of the photometry follows the recent
filter zeropoints published by STScI, $F606W = 26.420 - 2.5 {\rm log} (f)$
and $F814W = 25.536 - 2.5 {\rm log} (f)$ where $f$ represents the measured
counts per second.  These magnitudes can be converted to $(V,I)$
following the empirical transformations of
\citet{sir05}, namely $V = F606W + 0.236 (V-I)$ and
$I = F814W - 0.002 (V-I)$; we also use these in the later discussion.

The known Milky Way globular clusters (shown for comparison in the figure)
occupy a narrow range in colour, and a range in magnitude that is limited
to $F606W \la 24.5$ if projected to the distance of NGC 891.
However, we very conservatively 
define our ``best'', highest-confidence sample of GC candidates
to be the ones in the region $F606W < 23.7, (F606W-F814W) < 1.15$.
There are 16 such objects, which by definition have clearly survived
all of our rejection tests by scale size, ellipticity, and photometry.  
Although several others are in the same
blue color range and in the fainter magnitude range $23.7 < F606W < 24.5$,
close inspection of these fainter ones on the images shows that they are probably
made up of a wide mixture including small bulge or disk clusters,
faint and small background galaxies, and perhaps the occasional
classic GC.  In addition, if most of these actually were real GCs
then it would immediately imply that the globular cluster luminosity
function (GCLF) in NGC 891 would be very different from that of
the Milky Way, much more strongly weighted to the faint end.
For these reasons we do not rank them in the highest-confidence list.

We cannot immediately reject candidates that are redder than the 
Milky Way GC sequence
because individual objects \emph{may} be reddened by the heavy dust
lanes within NGC 891 itself.  However, clusters sitting behind 
significant amounts of dust extinction would be most likely to
fall within the range shown by the upper and lower reddening lines in the
figure. We therefore reject objects falling clearly \emph{below} this region,
i.e. those with $F606W > 24.5$ and $(F606W-F814W) \la 1.25$.  Again, close
inspection of the visual appearance of
these very faint objects shows that virtually all of them are
consistent with identification as small, distant background galaxies.

\begin{figure}
\includegraphics[angle=0,height=0.5\textwidth,width=0.5\textwidth]{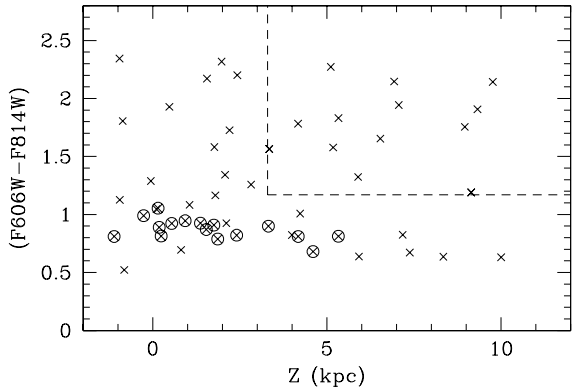}
\caption{Measured colour of candidates versus their distance above the plane
of NGC 891.
The 16 objects shown as large circled crosses
make up the ``best'' candidate GC sample
that are closest to the normal Milky Way cluster sequence in the previous figure.
All other, lower-quality candidates are plotted as crosses.
The objects to the upper right of the dashed lines are red ones
at large distances from the plane of the galaxy and can be rejected
(see text). Note that two objects just on the borderlines are included
in the final candidate list.}
\label{zcolor}
\end{figure}

In addition, we can reject very red objects that are located spatially well
away from the plane of NGC 891.  In Figure \ref{zcolor}, the 
colours for the remaining candidates are plotted against their projected distance
above the plane of the galaxy.  The large circled crosses show the 16 ``best'' objects that
fall closest to the normal GC sequence in the colour-magnitude diagram.
The candidates with $(F606W-F814W) \ga 1.17$ and 
$Z \ga 3$ kpc can be rejected since they are far redder than any known star 
clusters; note that no models of old simple stellar populations can account 
for such red colours, and these same objects are unlikely to be 
heavily reddened since there is no detectable trace of heavy dust extinction 
of any kind at such large distances from the plane. Using the H{\sc i} map 
of \citet{Oosterloo2007} to estimate the $E(B-V)$ reddening following the 
Galactic calibration of \citet{Rachford2008} indicates that although the 
correction is substantial near the galactic plane, the correction declines 
rapidly away from the plane, so that at the outer edge of the ACS survey 
region, the estimated internal extinction as derived from the H{\sc i} 
column density amounts to less than 0.001~mag.

Our final list of GC candidates, after all the culling steps described
above, consists of the 43 objects listed in Table 1. Successive columns 
give (1) a running ID number, (2) the ACS field on which it lies, (3,4) 
right ascension and declination (J2000) in decimal degree format as 
calculated from the astrometric parameters directly from the image headers,
(5,6) location $(x,y)$ (pixels) on the particular ACS field, 
(7-8) magnitudes, colours, and internal uncertainties, (9) half-light radius 
$r_h$ in arcseconds, (10) ellipse-fitted 
axial ratio $(b/a)$ as measured from \emph{ISHAPE}, and (11) any comment
on the object ranking or type; here, ``LSB'' means low surface brightness.

A definitive test of membership in NGC 891 would be direct measurement of 
radial velocity. The mean velocity of the galaxy is $v_r = 528$ km s$^{-1}$, 
thus its GCs should all have $300 {\rm km s}^{-1} \la v_r \la 800$ km s$^{-1}$. 
Because of our morphological selection criteria, 
the final candidate sample has no foreground stars, so the only remaining contaminants
can be faint background galaxies and perhaps a few massive open clusters in
the disk or bulge of NGC 891 itself. However, only 14 candidates are brighter 
than $V \simeq 23$, making velocity measurements challenging for all but 
these few.

\section{Discussion}

\begin{figure}
\includegraphics[angle=0,width=0.5\textwidth]{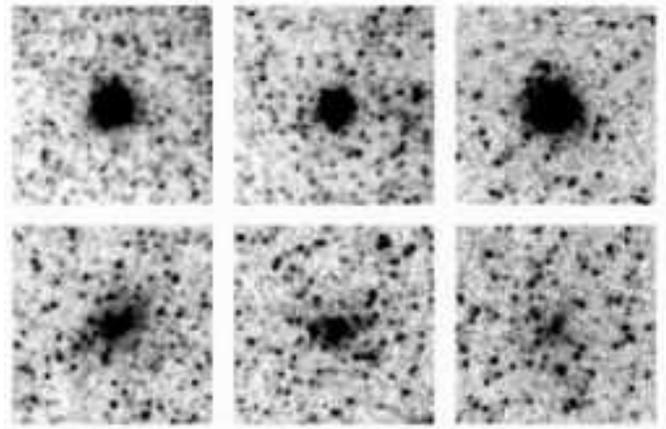}
\caption{Six of the candidate GCs in NGC 891, showing
their marginal resolution into stars.  From Table 1, these are 
G13, G24, and G34 (top row) and G22, G15, and G28 (bottom row).  
Each image here is 100 px across ($5''$).
Note the large numbers of field halo stars
within NGC 891 across each field.}
\label{samples}
\end{figure}

\begin{figure}
\includegraphics[angle=0,width=0.5\textwidth]{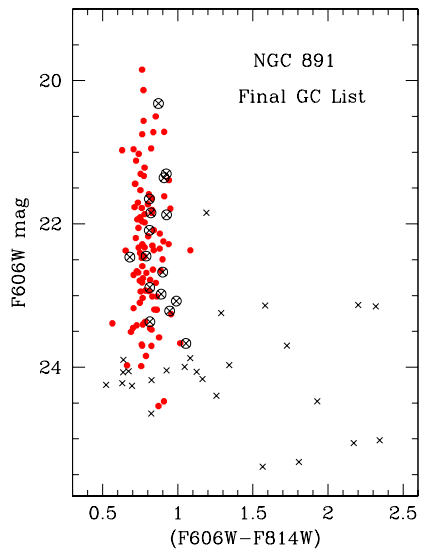}
\caption{The colour-magnitude diagram for the final sample of
43 GC candidates, compared with the Milky Way clusters (solid red
dots) as before.
The 16 best candidates are plotted as circled crosses and the 
remaining 27 lower-probability candidates as smaller crosses.
}
\label{cmdfinal}
\end{figure}

\begin{figure}
\includegraphics[angle=0,width=0.5\textwidth]{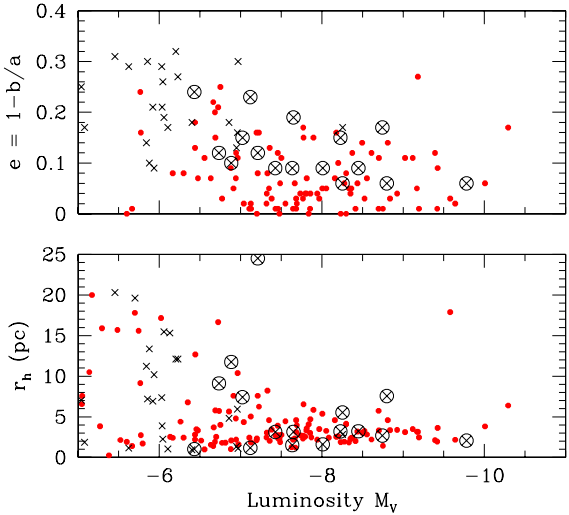}
\caption{Ellipticity (upper panel)  and half-light radius
(lower panel) plotted versus
cluster luminosity.  As before, the Milky Way clusters
are plotted as solid red dots, the 16 best NGC 891 candidates
as large circled crosses, and the 27 other candidates as
smaller crosses.
}
\label{aefinal}
\end{figure}

\begin{figure}
\includegraphics[angle=0,width=0.5\textwidth]{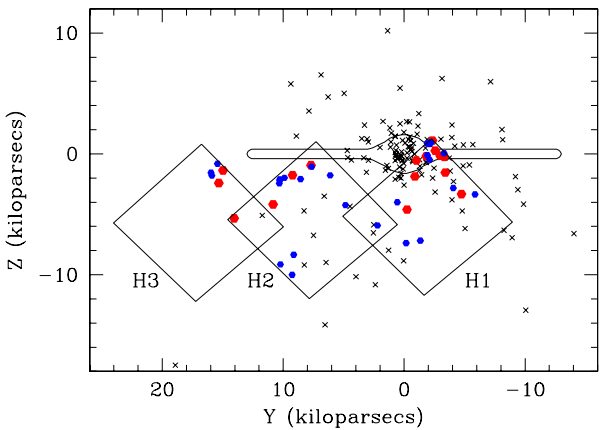}
\caption{Spatial locations of the selected GC candidates, superimposed
on the positions of the Milky Way clusters to the same scale.
A schematic outline of the Milky Way bulge and disk are drawn
in for scale comparison.
In this plane, the $X-$axis points roughly northward and $Z$
westward.
The 16 best GC candidates in NGC 891 are plotted as large red dots,
the remaining 27 as small blue dots, and the reference Milky Way
clusters as small crosses.}
\label{yz}
\end{figure}

\begin{figure}
\includegraphics[angle=0,width=0.5\textwidth]{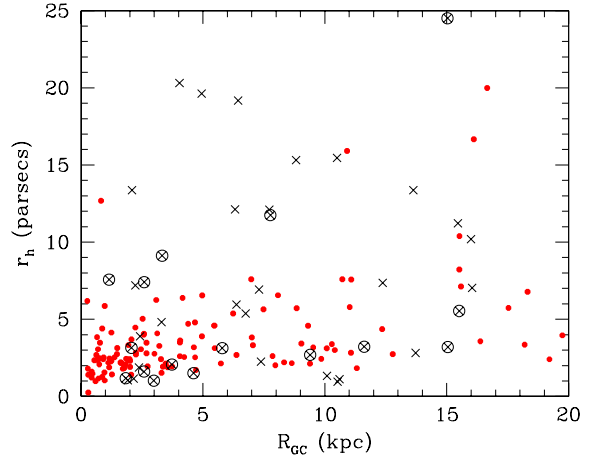}
\caption{Cluster effective radius $r_h$ plotted versus projected
galactocentric distance $R_{GC}$.  The large circled crosses
show the 16 best candidates from Table 1 while the
smaller crosses are for the remaining less certain candidates.
As in previous diagrams, the red solid dots show the same data for
the Milky Way clusters. Most of the largest candidates are also 
the faintest, thus their identification as true clusters is more
uncertain; see text.}
\label{rgc}
\end{figure}

After the many stages of weeding out individual contaminants, 
it is encouraging that we have numerous objects that do indeed 
resemble normal GCs closely even though the various culling stages were 
quite conservative. In Figure \ref{samples} we show thumbnail images of 
six of the candidates, demonstrating their partial resolution into 
stars.  In the top row, three of the brightest are shown, and in the
bottom row, three of the faintest.  As noted above, we regard the
identifications of the faintest ones as generally more uncertain,
and in some cases it may be more likely that they are actually
background galaxies.   However, we prefer to keep some
contaminants in the list rather than to miss a few real clusters.

More quantitative comparisons can be seen in Figures \ref{cmdfinal} and 
\ref{aefinal}. 
These two figures illustrate that our 16 best candidates have colours,   
magnitudes, half-light radii, and ellipticities all closely resembling
those of normal GCs in the Milky Way.
Full histogram comparisons between the two galaxies
to search for any finer differences
would be inconclusive given the small sample size of NGC 891 GCs.
The lack of NGC 891 candidates with \emph{formal} ellipticities
smaller than $\simeq 0.05$ (compared with the Milky Way, which has
many such clusters) cannot be assigned too much weight given
the internal fitting uncertainties in ISHAPE for these partially
resolved objects; but if real, it would indicate that the NGC 891
clusters are more similar in ellipticity to those in the Magellanic
Clouds or NGC 5128 \citep{har02,han94}.

The plot of $r_h$ versus luminosity shows hints of the
same trends in the Milky Way clusters, namely that the lower
envelope of points shows gradually increasing cluster scale
size with luminosity; and that the scatter in $r_h$ increases at
fainter luminosity.
The GC candidate with the biggest scale size, G40 at $r_h \simeq 24$ pc
(lying at the uppermost edge of the lower panel in Fig.~\ref{aefinal}),
is indeed an extended and bright object but does not show strong
traces of resolution into stars and \emph{may} be a background
galaxy.  Deeper images or radial velocity measurement would
make a definitive test.

The other 27 candidates mostly occupy the low-luminosity sides
of both graphs, and show larger dispersions in both $e$ and $r_h$
that also match the Milky Way spread quite well.  Unfortunately,
however, many of these candidates
are likely to be background galaxies as well.

As another comparison with the Milky Way, we show the spatial locations
of the NGC 891 clusters in Figure \ref{yz}, where $(Y,Z)$ in kiloparsec
units are the projected distance components parallel and perpendicular
to the plane of the galaxy.  The galactic center is at $(0,0)$.
The orientation of the figure is
chosen to match Figure 3 in \citet{Ibata2008}.
Here, as a schematic comparison,
the positions of the Milky Way GCs are shown projected on the
same plane, essentially viewing our galaxy as it would be seen from
outside, and looking inward along the line connecting the Sun to
the Galactic center.  This choice of two-dimensional
projection minimizes the effects of random distance errors to
individual Milky Way clusters, most of which lie toward the Galactic
center.  

Lastly, in Figure \ref{rgc} the distribution of cluster scale size
$r_h$ versus galactocentric distance is shown.   In this case the
Milky Way data (small dots) are for \emph{projected} Galactocentric
distance $R_{GC} \equiv \sqrt{Y^2+Z^2}$ to make the graph
for the two galaxies strictly
comparable.  In the Milky Way we see the well known trend for
the mean $r_h$ to increase gradually with $R_{GC}$ \citep[e.g.][]{vdb91,jor05}
and for the range in $r_h$ to increase at larger $R_{GC}$ as well.
The NGC 891 data hint barely at the same increase amidst the
large scatter.  Five of the NGC 891 clusters have radii that fall
clearly above the Milky Way distribution, however; these are
G07, G08, G13, G19, and G40.  If at least some of these five
are not contaminants, we can speculate that they may have
originated in dwarf satellite galaxies within which GCs have
characteristically larger radii \citep[see][for a recent discussion]{dac09}.
Such an origin would mesh well with our analysis of the
NGC 891 stellar halo \citep{Ibata2008} which shows significant
substructure and indicates that at least some part of it has
been accreted from satellites.  Without more material than our
presently rather slim GC sample to base it on, it is risky to carry such
discussion further.

Field H1 is closest
to the galaxy center and bears out our expectations that
it should contain the most GCs of the three fields 
(see Fig.~\ref{yz}).  Furthermore,
any GCs particularly in H1 that happen to
lie behind the heavy dust
lanes there may well be escaping detection, so by symmetry
we might then expect that the true total GC population in H1 is
as much as twice as large as the 10 good candidates and 9 lesser-quality
ones we actually see.
Combining all the preceding arguments, we estimate \emph{very roughly}
that our three observed fields may contain $25 \pm 5$ high-quality GC
candidates over all magnitudes
after correction for contamination and for losses due to extinction.

Estimating the total GC population within the entire galaxy  
requires a large extrapolation from the severely limited area coverage
of our survey.  Our three ACS fields cover the region to the east
side of the northern disk, so they can take in at most $\sim 1/4$ of
the entire population.  We therefore suggest, albeit tentatively, that 
the total should be $N_T \ga 100$ and perhaps as large as $\sim 140$.
Alternately, we can compare the number of cluster candidates in our
three fields with the number of Milky Way clusters that fall within
the same area when projected onto the same scale (see Fig.~\ref{yz}).
In the Figure, 30 to 35 Milky Way clusters fall within the marked-out
areas of H1/2/3, compared with anywhere from 16 to 43 candidates in
our defined list for NGC 891.  Since the total cataloged population of Milky 
Way clusters is approximately 150, these ratios suggest that the
total population of NGC 891 globular clusters is in the range
of $\sim 80$ to 200, in good agreement with the previous estimate.
The resulting specific frequency \citep{vdb82} is then $S_N \sim 0.3 - 0.5$, 
if we adopt a galaxy luminosity $V_T^0 = 8.82$ from the NED database.  
This specific frequency would put NGC 891 reasonably in the range of most 
known disk galaxies similar to the Milky Way \citep[e.g.][]{har01}.

In summary, our admittedly restricted look at the globular cluster
system of NGC 891 leads us to conclude that it resembles that of
the Milky Way rather closely in approximate total numbers, spatial
distribution, and structural properties of the individual clusters.
Our data clearly cover only a very limited region of the NGC 891
halo, and a more comprehensive targetted search should be able to
increase the GC sample by a factor of five or more.

\begin{table*}
\caption{Candidate Globular Clusters in NGC 891}
\label{list_gc_candidates}
\begin{tabular}{lllllllllll}
\hline
ID & Field &  RA      &     Dec    & $x$ (px)& $y$ (px)& $F606W$            & $(F606W-F814W)$   &$r_h$ (arcsec) & $(b/a)$ & Comment\\
\hline
G01 & H1 &  35.624222 &  42.338615 &  5480.3 &  3664.0 & 22.089 $\pm$  0.007 &  0.811 $\pm$  0.013 &   0.0342 &   0.91 & best \\
G02 & H1 &  35.625717 &  42.338867 &  5400.6 &  3682.1 & 24.065 $\pm$  0.041 &  1.126 $\pm$  0.070 &   0.0828 &   0.79 & bulge \\
G03 & H1 &  35.625805 &  42.339149 &  5396.0 &  3702.4 & 25.020 $\pm$  0.086 &  2.344 $\pm$  0.093 &   0.0397 &   0.83 & bulge \\
G04 & H1 &  35.627125 &  42.339451 &  5325.7 &  3724.1 & 24.245 $\pm$  0.111 &  0.522 $\pm$  0.216 &   0.1530 &   0.70 & bulge \\
G05 & H1 &  35.627289 &  42.340481 &  5316.9 &  3798.3 & 25.323 $\pm$  0.291 &  1.806 $\pm$  0.319 &   0.2844 &   0.84 & bulge \\
G06 & H1 &  35.629326 &  42.331100 &  5209.0 &  3122.7 & 23.243 $\pm$  0.035 &  1.289 $\pm$  0.048 &   0.1025 &   0.82 & disk \\
G07 & H1 &  35.629837 &  42.335514 &  5181.5 &  3440.6 & 23.077 $\pm$  0.031 &  0.990 $\pm$  0.051 &   0.1576 &   0.85 & best \\
G08 & H1 &  35.631474 &  42.330360 &  5094.6 &  3069.6 & 23.366 $\pm$  0.029 &  0.814 $\pm$  0.050 &   0.1939 &   0.88 & best \\
G09 & H1 &  35.631748 &  42.332382 &  5079.9 &  3215.0 & 23.669 $\pm$  0.053 &  1.054 $\pm$  0.084 &   0.0217 &   0.76 & best \\
G10 & H1 &  35.634480 &  42.338356 &  4934.3 &  3645.2 & 23.996 $\pm$  0.035 &  1.046 $\pm$  0.068 &   0.0222 &   0.83 & disk \\
G11 & H1 &  35.635334 &  42.338711 &  4888.8 &  3670.6 & 22.981 $\pm$  0.012 &  0.889 $\pm$  0.031 &   0.0250 &   0.77 & best \\
G12 & H1 &  35.636715 &  42.336533 &  4815.4 &  3513.9 & 24.476 $\pm$  0.143 &  1.928 $\pm$  0.155 &   0.0246 &   0.71 & disk \\
G13 & H1 &  35.640335 &  42.342419 &  4622.6 &  3937.4 & 21.301 $\pm$  0.006 &  0.923 $\pm$  0.009 &   0.1609 &   0.94 & best \\
G14 & H1 &  35.640930 &  42.327168 &  4591.4 &  2839.4 & 20.320 $\pm$  0.002 &  0.871 $\pm$  0.003 &   0.0440 &   0.94 & best \\
G15 & H1 &  35.647358 &  42.309727 &  4249.5 &  1583.5 & 25.388 $\pm$  0.034 &  1.565 $\pm$  0.039 &   0.1142 &   0.88 & LSB \\
G16 & H1 &  35.648632 &  42.320786 &  4181.6 &  2379.6 & 24.399 $\pm$  0.018 &  1.258 $\pm$  0.023 &   0.4173 &   0.36 & galaxy? \\
G17 & H1 &  35.650364 &  42.315956 &  4089.4 &  2031.9 & 22.673 $\pm$  0.005 &  0.900 $\pm$  0.007 &   0.0666 &   0.91 & best \\
G18 & H1 &  35.650578 &  42.340332 &  4077.4 &  3786.9 & 22.450 $\pm$  0.006 &  0.789 $\pm$  0.009 &   0.0669 &   0.81 & best \\
G19 & H2 &  35.668228 &  42.389709 &  4683.5 &  3592.4 & 23.215 $\pm$  0.014 &  0.946 $\pm$  0.021 &   0.2499 &   0.90 & best \\
G20 & H2 &  35.668983 &  42.389130 &  4643.3 &  3550.5 & 23.872 $\pm$  0.020 &  1.083 $\pm$  0.032 &   0.2574 &   0.73 & disk \\
G21 & H2 &  35.669926 &  42.379181 &  4593.4 &  2834.3 & 23.140 $\pm$  0.007 &  1.581 $\pm$  0.009 &   0.1266 &   0.84 & galaxy?\\
G22 & H1 &  35.670647 &  42.343815 &  3009.4 &  4037.5 & 24.646 $\pm$  0.023 &  0.823 $\pm$  0.033 &   0.4319 &   0.69 & LSB \\
G23 & H1 &  35.672882 &  42.338108 &  2890.5 &  3626.5 & 22.464 $\pm$  0.004 &  0.681 $\pm$  0.006 &   0.0321 &   0.91 & best \\
G24 & H2 &  35.678722 &  42.396378 &  4125.4 &  4072.4 & 21.355 $\pm$  0.003 &  0.909 $\pm$  0.004 &   0.0573 &   0.83 & best \\
G25 & H2 &  35.679241 &  42.392036 &  4097.8 &  3759.7 & 23.972 $\pm$  0.014 &  1.342 $\pm$  0.018 &   0.3257 &   0.43 & galaxy? \\
G26 & H2 &  35.682297 &  42.399582 &  3935.3 &  4302.9 & 23.150 $\pm$  0.006 &  2.318 $\pm$  0.007 &   0.0285 &   0.87 & cluster?\\
G27 & H2 &  35.684418 &  42.401455 &  3822.5 &  4437.9 & 24.043 $\pm$  0.011 &  0.925 $\pm$  0.015 &   0.3288 &   0.81 & cluster?\\
G28 & H2 &  35.684708 &  42.366974 &  3807.5 &  1955.0 & 25.844 $\pm$  0.044 &  1.008 $\pm$  0.059 &   0.4078 &   0.36 & LSB \\
G29 & H2 &  35.685135 &  42.401295 &  3784.2 &  4426.2 & 23.698 $\pm$  0.008 &  1.727 $\pm$  0.009 &   0.0202 &   0.82 & star? \\
G30 & H2 &  35.686897 &  42.400967 &  3690.6 &  4402.6 & 23.132 $\pm$  0.006 &  2.201 $\pm$  0.007 &   0.0238 &   0.70 & star? \\
G31 & H1 &  35.688915 &  42.326588 &  2037.0 &  2797.1 & 24.181 $\pm$  0.010 &  0.825 $\pm$  0.015 &   0.1472 &   0.79 & galaxy? \\
G32 & H3 &  35.689495 &  42.432602 &  5072.7 &  2963.6 & 24.260 $\pm$  0.015 &  0.695 $\pm$  0.024 &   0.2386 &   0.86 & galaxy? \\
G33 & H2 &  35.689682 &  42.348801 &  3542.9 &   646.6 & 23.897 $\pm$  0.009 &  0.638 $\pm$  0.013 &   0.2577 &   0.68 & galaxy? \\
G34 & H3 &  35.692303 &  42.428921 &  4923.5 &  2698.4 & 21.872 $\pm$  0.003 &  0.925 $\pm$  0.004 &   0.0680 &   0.85 & best \\
G35 & H1 &  35.693783 &  42.332630 &  1778.0 &  3232.2 & 24.057 $\pm$  0.010 &  0.672 $\pm$  0.015 &   0.0479 &   0.74 & galaxy? \\
G36 & H3 &  35.696560 &  42.433994 &  4697.0 &  3063.5 & 25.059 $\pm$  0.022 &  2.172 $\pm$  0.024 &   0.1496 &   0.75 & galaxy? \\
G37 & H3 &  35.698162 &  42.433067 &  4612.1 &  2996.9 & 24.165 $\pm$  0.011 &  1.165 $\pm$  0.014 &   0.2167 &   0.91 & galaxy? \\
G38 & H3 &  35.701077 &  42.428608 &  4457.4 &  2675.5 & 21.846 $\pm$  0.003 &  0.821 $\pm$  0.004 &   0.1177 &   0.94 & best \\
G39 & H2 &  35.701416 &  42.400101 &  2918.6 &  4340.1 & 21.651 $\pm$  0.003 &  0.811 $\pm$  0.004 &   0.0685 &   0.91 & best \\
G40 & H2 &  35.719254 &  42.415409 &  1970.6 &  5442.5 & 22.888 $\pm$  0.005 &  0.813 $\pm$  0.007 &   0.5212 &   0.88 & best \\
G41 & H2 &  35.727627 &  42.381878 &  1524.6 &  3028.4 & 24.070 $\pm$  0.010 &  0.636 $\pm$  0.015 &   0.1566 &   0.71 & galaxy? \\
G42 & H2 &  35.736752 &  42.386242 &  1039.4 &  3342.8 & 21.844 $\pm$  0.003 &  1.190 $\pm$  0.004 &   0.0599 &   0.83 & galaxy? \\
G43 & H2 &  35.740395 &  42.379120 &   845.3 &  2830.1 & 24.223 $\pm$  0.010 &  0.631 $\pm$  0.016 &   0.2842 &   0.90 & galaxy? \\
\end{tabular}
\end{table*}

\section*{Acknowledgements}

This work was supported by the Natural Sciences and Engineering
Research Council of Canada through research grants to WEH.

\bsp

\label{lastpage}

\end{document}